\documentclass[prd,twocolumn,showpacs,floatfix,amsmath,nofootinbib,amssymb,floatfix]{revtex4}
\usepackage{graphicx,color,dcolumn,booktabs,bm,multirow}
\usepackage{longtable,lscape}
\usepackage{txfonts}
\usepackage{overpic}
\usepackage{amssymb}
\usepackage{indentfirst}
\usepackage{feynmf}   
\usepackage{slashed}  
\usepackage{cases}
\usepackage{color}
\usepackage{multirow}
\usepackage{epstopdf}
\usepackage{graphicx,color,dcolumn,booktabs,bm}
\usepackage{epstopdf}
\usepackage{ulem}

\usepackage[colorlinks,
                     citecolor=green,
                     anchorcolor=red,
                     menucolor=red,
                     linkcolor=blue,
                     filecolor=red,
                     runcolor=red,
                     urlcolor=blue,
                     frenchlinks=red]{hyperref}

\def\lrpartial{\buildrel\leftrightarrow\over\partial}

\begin{document}

\title{Understanding the $\eta_c\rho$ decay mode of $Z_c^{(\prime)}$ via {the triangle loop mechanism} }
\author{Cheng-Jian Xiao$^{1}$}\email{xiaocj@ihep.ac.cn}
\author{Dian-Yong Chen$^{2}$\footnote{Corresponding author}}\email{chendy@seu.edu.cn}
\author{Yu-Bing Dong$^1$}\email{dongyb@ihep.ac.cn}
\author{Wei Zuo$^3$}\email{zuowei@impcas.ac.cn}
\author{Takayuki Matsuki$^{4,5}$}\email{matsuki@tokyo-kasei.ac.jp}
\affiliation{
$^1$Institute of High Energy Physics, Chinese Academy of Sciences,
    Beijing 100049, People's Republic of China\\
$^2$ School of Physics, Southeast University,
    Nanjing 210094, People's Republic of China\\
$^3$Institute of Modern Physics, Chinese Academy of Sciences,
    Lanzhou 730000, People's Republic of China \\
$^4$Tokyo Kasei University, 1-18-1 Kaga, Itabashi, Tokyo 173-8602, Japan \\
$^5$Theoretical Research Division, Nishina Center, RIKEN, Wako, Saitama 351-0198, Japan}

\date{\today}
\begin{abstract}
{Recently, the BESIII Collaboration reported a new measurement of the $\eta_c \rho$ decay mode of $Z_c^{(\prime)}$, which motivated us to study the inner structure of $Z_c^{(\prime)}$  via investigating the hidden charm decays of these two $Z_c$ states. We consider the {triangle loop mechanism} contribution in the hidden charm decays of $Z_c^{(\prime)}$. Our estimations indicate that the triangle loop mechanism plays an important role in the decays of the $Z_c^{(\prime)}$, where our results are in agreement with the experimental observations  in a reasonable parameter range. Furthermore, we point out that the $Z_c^{(\prime)}$ can be interpreted as the hadronic molecules, while the tetraquark scenario is less favored.}
\end{abstract}


\maketitle

\section{Introduction}\label{sec:1}

In 2013, a new charged charmoniumlike state $Z_c(3900)$ in the $\pi^{\pm} J/\psi$ invariant mass spectrum was reported by the BESIII and Belle Collaboration in the $e^+ e^- \to \pi^+ \pi^- J/\psi$  process at 4.60 GeV \cite{Ablikim:2013mio, Liu:2013dau}. The statistical significance of $Z_c(3900)$ is more than 8$\sigma$ and the measured resonance parameters are $m=(3899\pm3.6\pm4.9)$ MeV and $\Gamma=(46\pm10\pm20)$ MeV \cite{Ablikim:2013mio}.  Later, this resonance was confirmed by the CLEO-c  in the same process but at $\sqrt s=4.17$\,GeV \cite{Xiao:2013iha}, and the neutral partner was also observed for the first time.  In the same year, the BESIII analyzed the $e^+e^-\to\pi^+\pi^-h_c$ process just after the observation of $Z_c(3900)$ and they found a similar charmoniumlike state named $Z_c(4020)^\pm$ in the $\pi^\pm h_c$ invariant mass spectrum \cite{Ablikim:2013wzq}. The measured mass and width are $(4022.9\pm0.8\pm2.7)$\,MeV and $(7.9\pm2.7\pm2.6)$\,MeV, respectively, where the significance is more than 8.9$\sigma$\cite{Ablikim:2013wzq}. In the process $e^+e^-\to \pi^0\pi^0h_c$ at $\sqrt s=4.23,\ 4.26,\ 4.36$\,GeV, the neutral  $Z_c(4020)^0$ was also observed by the BESIII Collaboration\cite{Ablikim:2014dxl} and therefore, $Z_c(4020)$ is an isovector state.

In the open charm process, both $Z_c(3900)$ and $Z_c(4020)$ have been observed by the BESIII Collaboration.  In 2014, $Z_c(3900)$ was observed in the $\bar D^\ast D$ invariant mass spectrum of the $e^+e^-\to \pi^\pm(\bar D^\ast D)^\mp$ process \cite{Ablikim:2013xfr}.  Performing the partial wave analysis, the $J^P$ quantum number of the $Z_c(3900)$ is determined to be $1^+$ \cite{Collaboration:2017njt}. Similarly, $Z_c(4020)$ was observed in the $D^\ast\bar D^\ast$ invariant mass spectrum of the $e^+e^- \to \pi^{\pm} (D^\ast \bar{D})^{\mp}$ process by the BESIII Collaboration \cite{Ablikim:2013emm}. It is interesting to notice that $Z_c(4020)$ is observed only in the $D^\ast \bar{D}^\ast$ mass spectrum, while in the $D^\ast \bar{D}$ mode, it is not observed. As for $Z_c(3900)$, it can only decay into $D^\ast\bar D$, because the $D^\ast\bar D^\ast$ mode is forbidden due to the limited phase space.

Hereafter, we adopt $Z_c$ and $Z_c^\prime$ to refer to $Z_c(3900)$ and $Z_c(4020)$, respectively. Comparing to other charmoniumlike $X,\ Y,\ Z$ states, these two charged $Z_c$ and $Z_c^\prime$ are the first confirmed charged charmoniumlike states and  they contain at least four constituent quarks. Therefore, they are ideal candidates of the tetraquark state. The author in Ref.~\cite{Chen:2010ze} predicted a $1^{++}$ $qc\bar q\bar c$ state around 4.0\,GeV before the observation of the two $Z_c$ states.  Assuming that the $Z_c$ is a tetraquark, its mass can be calculated in Refs. \cite{Faccini:2013lda,Braaten:2013boa,Goerke:2016hxf,Qiao:2013raa,Wang:2013vex,Deng:2014gqa,Agaev:2016dev}. The decays of $Z_c$ in the tetraquark assumption were investigated in Refs.~\cite{Agaev:2016dev,Dias:2013xfa,Wang:2017lot,Esposito:2014hsa}. The tetraquark picture was also proposed to explain the $Z_c^\prime$ \cite{Esposito:2014hsa,Deng:2014gqa,Qiao:2013dda,Wang:2013exa}.

In addition, it was noticed that the measured masses of $Z_c$ and $Z_c^\prime$ locate near the threshold of $D^\ast\bar D$ and $D^\ast\bar D^\ast$, respectively, which indicates that $Z_c$ and $Z_c^\prime$ could be good candidates of hadronic molecules composed of $D^\ast \bar{D}+h.c$ and $D^\ast \bar{D}^\ast$, respectively. Considering the potential caused by the one-boson exchange processes,  the $D^\ast \bar D$ and $D^\ast\bar D^\ast$ molecules were predicted in Ref.~\cite{Sun:2012zzd}. The authors in Ref. \cite{Aceti:2014uea} studied the $D^\ast \bar D$ interaction and they found a bound state around 3869$-$3875\,MeV, which is consistent with the mass of $Z_c$~\cite{Aceti:2014kja}.  In Ref.~\cite{He:2015mja},  $Z_c$ was interpreted as a resonance resulted from $D^\ast \bar D$ interaction. In Refs.~\cite{Esposito:2014hsa,Dong:2013iqa,Wilbring:2013cha,Goerke:2016hxf,Khemchandani:2013iwa,Guo:2013sya,Ke:2013gia,Chen:2013omd,Wang:2013cya,Li:2014pfa}, the $Z_c^{(\prime)}$ were considered as the $D^\ast\bar D^{(\ast)}$ molecular states and theirs decays were investigated. The intermediate meson loops model was also used to investigate the decays of the $Z_c^{(\prime)}$\cite{Li:2013xia}.In addition, the threshold effects, such as the cusp effect and the initial single pion emission mechanism, were proposed as the source of $Z_c^{(\prime)}$~\cite{Swanson:2014tra,Liu:2013vfa,Ikeda:2016zwx,Voloshin:2013dpa,Chen:2011xk,Chen:2013coa,Chen:2013wca}.

{To date, the inner structure of these two charged charmoniumlike states has been unknown, and more efforts are necessary to reveal their nature. On the theoretical side, the study on the decay properties of the $Z_c$ and $Z_c^\prime$ is important. By comparing the theoretical estimations with the experimental measured decay properties, the nature of $Z_c$ and $Z_c^\prime$ might be no longer be ambiguous. In other words, the more decay processes are experimentally measured, the more it helps us to distinguish the nature of $Z_c$ and $Z_c^\prime$. Therefore, the new measurements of the $Z_c^{(\prime)} \to \eta_c\rho$ processes from the BESIII Collaboration are very useful.  The ratios of the partial widths of $Z_c^{(\prime)} \to \eta_c\rho$ and $J/\psi \pi$ at $\sqrt s=4.23$\,GeV are measured to be \cite{Yuan:2018inv}}
{\begin{eqnarray}
R&\equiv& \frac{\Gamma(Z_c\to\eta_c\rho)}{\Gamma(Z_c\to J/\psi\pi)}=2.1\pm0.8,\label{eq:ratio-etacrho-jpsipi}\\
R^\prime &\equiv& \frac{\Gamma(Z_c^\prime\to\eta_c\rho)}{\Gamma(Z_c^\prime\to h_c\pi)}<1.9\label{eq:ratio-etacrho-hcpi}.
\end{eqnarray}
This experimentally measured ratio $R^{(\prime)}$  could be a good tool for detecting the inner structure of the $Z_c^{(\prime)}$.

Before the recent measurements of the $\eta_c\rho$ mode of $Z_c$ and $Z_c^{\prime}$, the BESIII Collaboration had already measured the open charm decay modes and found that the $Z_c^{(\prime)}$ dominantly decays into a pair of charmed mesons\cite{Ablikim:2013xfr,Ablikim:2013emm}. Therefore, the $Z_c$ and $Z_c^{\prime}$ can decay into the hidden charm final states via an intermediate charmed meson loop, where the pair of charmed mesons could connect the hidden charm final states and the initial $Z_c^{(\prime)}$ by exchanging a proper charmed meson. Such a triangle loop mechanism or meson loop mechanism has been widely employed to investigate the hidden charm and bottom decays of higher charmonia and bottomonia  \cite{Meng:2007tk,Meng:2008bq}. One may wonder if a study of the decays of $Z_c^{(\prime)}$ to hidden charm mesons, proceeding through intermediate loops of the open charm meson, is based on the inner structure of the decaying particle. For instance, can such an investigation help us to distinguish the nature of $Z_c^{(\prime)}$, whether it is a hadronic molecule or a tetraquark? It is important to realize that the decay mechanism is different for the decay of a hadronic molecule and for a tetraquark state. In the case of a hadronic molecule, the $Z_c^{(\prime)}$ decays to the hidden charm final states via their charmed meson components and the loop mechanism is the primary decay mechanism. As for the tetraquark state, it can directly decay into the hidden charm state via the rearrangement of the four constituent quarks, which should be the major contribution to the hidden charm decays compared to the triangle loop mechanism. Thus, in this sense, the role the triangle loop mechanism takes in the hidden charm decays of $Z_c^{(\prime)}$ could be used to distinguish the molecular and tetraquark scenarios.

In the present work, the triangle loop mechanism is applied to estimate the recent measurements of the $Z_c^{(\prime)}\to\eta_c\rho$ processes from the BESIII Collaboration. To directly compare our results with the experimental measured ratios, we also estimate the widths of $Z_c\to J/\psi \pi $ and $Z_c^\prime \to h_c \pi$ with the same mechanism. By comparing our estimations with the experimental decay behaviors of $Z_c$ and $Z_c^\prime$, we can better understand the nature of these two $Z_c$ states.
}

This work is organized as follows. After the Introduction, we present the amplitudes of hidden charm decays of $Z_c$ and $Z_c^\prime$ in Sec.~\ref{sec:2}, and the numerical results and discussions are given in Sec.~\ref{sec:3}. Sec.~\ref{sec:4} is devoted to a short summary.

\section{The hidden charm decays of $Z_c(3900)$ and $Z_c(4020)$}\label{sec:2}

In the present work, we apply the  effective Lagrangian approach to estimate the hidden charm decays of $Z_c^{(\prime)}$. The effective Lagrangian describing the  $Z_c^{(\prime)} $ and $D^{\ast} D^{(\ast)}$ interactions are
\begin{eqnarray}
\mathcal{L}_{Z_c D^\ast D}&=& g_{Z_c}Z_c^\mu (D^+\bar D^{\ast0}_\mu+ D^{\ast+}_\mu \bar D^0),\nonumber\\
\mathcal{L}_{Z_c^\prime D^\ast D^\ast}&=& ig_{Z_c^\prime} \epsilon^{\mu\nu\alpha\beta}\partial_\mu Z_{c\nu}^\prime D^{\ast+}_\alpha\bar D^{\ast0}_\beta.\label{eq:LagZc}
\end{eqnarray}
In the heavy quark limit, one can construct the  effective Lagrangian of charmonia and charmed meson pair couplings, which are \cite{Casalbuoni:1996pg,Colangelo:2003sa,Chen:2015igx}
\begin{eqnarray}
\mathcal{L}_{\psi D^{(\ast)}D^{(\ast)}}&=&
    -ig_{\psi DD}\psi_{\mu}
    (\partial^\mu \mathcal {DD}^\dagger-\mathcal D\partial^\mu \mathcal D^\dagger)\nonumber\\
&&+
    g_{\psi D^\ast D}\epsilon^{\mu\nu\alpha\beta}
    \partial_{\mu}\psi_{\nu}(\mathcal D^{\ast}_\alpha \lrpartial_\beta \mathcal D^\dagger-\mathcal D\lrpartial_\beta \mathcal D^{\ast\dagger}_{\alpha})\nonumber\\
&&+
    ig_{\psi D^\ast D^\ast}\psi^\mu
    (\mathcal D^{\ast}_\nu\lrpartial\!{}^\nu \mathcal D^{\ast\dagger}_\mu+ \mathcal D^{\ast}_\mu \lrpartial\!{}^\nu \mathcal D^{\ast\dagger}_\nu\nonumber\\
&&-\mathcal D^{\ast}_\nu\lrpartial_\mu \mathcal D^{\ast\dagger\nu}),\\
\mathcal L_{\eta_c D^{\ast}D^{(\ast)}}&=&
    -ig_{\eta_c \mathcal D^\ast \mathcal D}\eta_c(\mathcal D\lrpartial_\mu  {\mathcal D}^{\ast\dagger\mu}+\mathcal D^{\ast\mu}\lrpartial_\mu {\mathcal D^\dagger})\nonumber\\
  &&-g_{\eta_c D^\ast D^\ast}\epsilon^{\mu\nu\alpha\beta}\partial_\mu\eta \mathcal D^{\ast \nu}\lrpartial\!{}^\alpha {\mathcal D}^{\ast\dagger\beta},\\
\mathcal{L}_{h_c D^\ast D^{(\ast)}}&=&g_{h_c \mathcal D^\ast \mathcal D}h_c^\mu
    ( \mathcal D {\mathcal D}^{\ast\dagger}_\mu +\mathcal D^\ast_\mu {\mathcal D^\dagger})\nonumber\\
&&+ig_{h_cD^\ast D^\ast}\epsilon^{\mu\nu\alpha\beta}
    \partial_\mu h_{c\nu}\mathcal D^\ast_\alpha {\mathcal D}^{\ast\dagger}_\beta.
\end{eqnarray}
Considering the heavy quark limit and chiral symmetry, the effective Lagrangian related to the light mesons are \cite{Chen:2014sra}
\begin{eqnarray}
\mathcal{L}_{D^\ast D^{(\ast)}P}&=&-ig_{D^\ast DP}( {\mathcal D^\dagger}\partial^{\mu}
    \mathcal P \mathcal D^{\ast}_\mu - {\mathcal D}^{\ast\dagger\mu}\partial_\mu\mathcal P \mathcal D) \nonumber\\
&&+\frac{1}{2}g_{D^\ast D^\ast P}\epsilon^{\mu\nu\alpha\beta}
      {\mathcal D}^{\ast\dagger}_\mu\partial_{\nu}\mathcal P \lrpartial_{\alpha}\mathcal D^{\ast}_\beta,\\
\mathcal{L}_{D^{(\ast)}D^{(\ast)}V}&=&-ig_{DDV}\mathcal D^\dagger_i
   \lrpartial\!{}_{\mu}\mathcal D^j(\mathcal V_\mu)^i_j-2f_{D^\ast DV}\epsilon^{\mu\nu\alpha\beta}\nonumber\\
&&\times(\partial_{\mu}\mathcal V_\nu)^i_j
     (\mathcal D_i^\dagger\lrpartial_{\alpha}\mathcal D^{\ast j}_\beta
     -\mathcal D^{\ast\dagger}_{\beta i}\lrpartial_{\alpha}\mathcal D^j)\nonumber\\
&&+ig_{D^*D^*V}\mathcal D^{\ast\dagger\nu}_i\lrpartial{}_{\mu}
     \mathcal D_{\nu}^{\ast j}(\mathcal V^{\mu})^i_j\nonumber\\
&&+4if_{D^*D^*V}\mathcal D^{\ast\dagger{\mu}}_i(\partial_{\mu}\mathcal V^\nu-
     \partial^{\nu}\mathcal V_{\mu})^i_j \mathcal D_{\nu }^{\ast j},\label{eq:la-bbv}
\end{eqnarray}
where the ${\mathcal{D}}^{(\ast)\dagger}$=$(\bar D^{(\ast)0},D^{(\ast)-},D_s^{(\ast)-})$ is the charmed meson triplets. $\mathcal{P}$ and $\mathcal{V}$ are matrices of  pseudoscalar and vector mesons. Their explicit forms are
\begin{eqnarray}\label{eq:P-matrix}
\mathcal{P} &=&
{\small\left(\begin{array}{ccc}
\frac{\pi^{0}}{\sqrt 2}+\alpha\eta+\beta\eta\prime &\pi^{+} &K^{+}\\
\pi^{-} &-\frac{\pi^{0}}{\sqrt2}+\alpha\eta+\beta\eta \prime&K^{0}\\
K^{-} &\bar K^{0} &\gamma\eta+\delta\eta\prime
\end{array}\right)},\nonumber\\ \\
\hspace*{\fill}\mathcal{V} &=&
{\small\left(
\begin{matrix}
\frac{1}{\sqrt2}(\rho^0+\omega)&\rho^{+}&K^{*+}\\
\rho^{-}&\frac{1}{\sqrt2}(-\rho^{0}+\omega)&K^{*0}\\
K^{*-}&\bar K^{*0}&\phi
\end{matrix}
\right)}\hspace*{\fill}, \label{eq:V-matrix-ideal}
\end{eqnarray}
where the parameters related to the mixing angle are defined as
\begin{eqnarray}
&&\alpha={\cos{\theta}-{\sqrt2}\sin{\theta}\over\sqrt6},
\
\beta=\frac{\sin{\theta}+{\sqrt2}\cos{\theta}}{\sqrt6},\\
&&\gamma=\frac{-2\cos{\theta}-{\sqrt2}\sin{\theta}}{\sqrt6},\
\delta=\frac{-2\sin{\theta}+{\sqrt2}\cos{\theta}}{\sqrt6},
\end{eqnarray}
and the  mixing angle is $\theta=-19.1^\circ$.

\subsection{The hidden charm decays of $Z_c$}
\begin{figure}[hbt!]
  \begin{tabular}{cccc}
\includegraphics[scale=0.35]{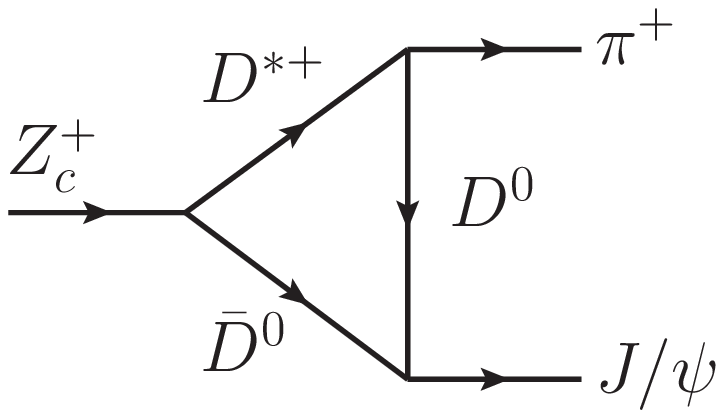}
&\includegraphics[scale=0.35]{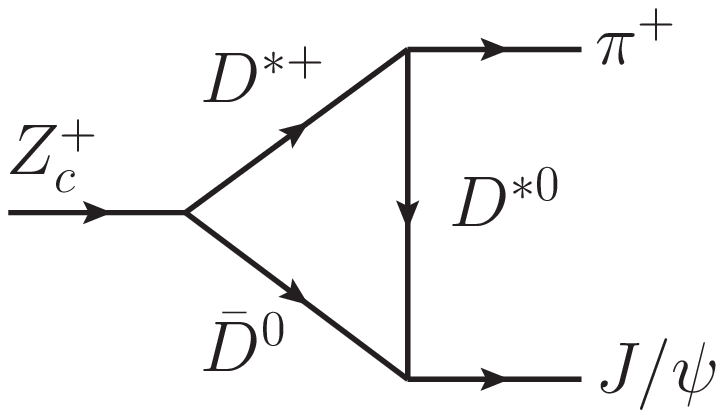}
&\includegraphics[scale=0.35]{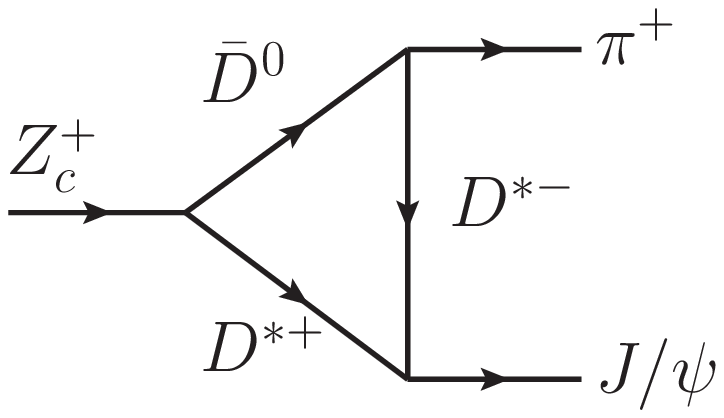}\\
(a) &(b)    &(c)\\ \\
\includegraphics[scale=0.35]{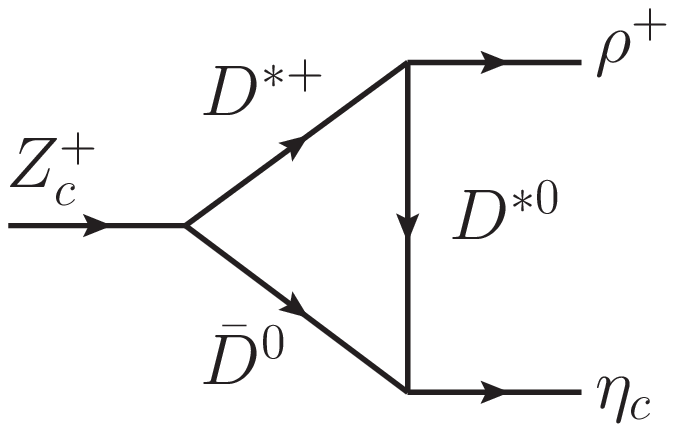}
&\includegraphics[scale=0.35]{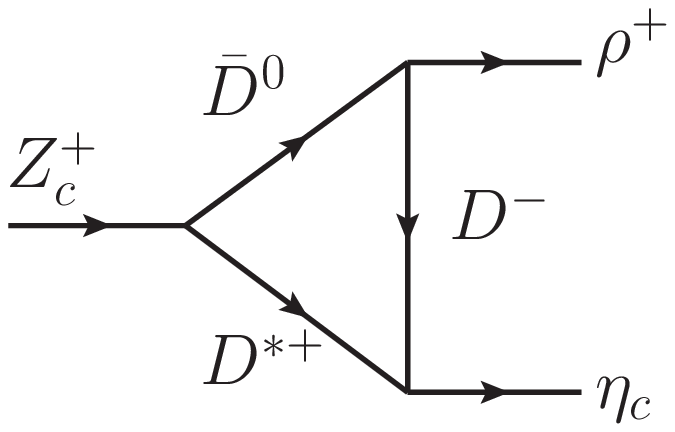}
&\includegraphics[scale=0.35]{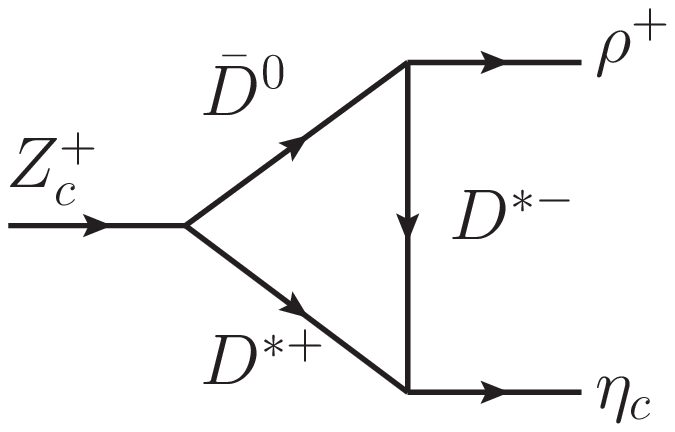}\\
(d) & (e) & (f)
  \end{tabular}
\caption{Typical diagrams contributing to the processes $Z_c\to J/\psi\pi$ (diagrams (a)-(c)) and $\eta_c \rho$ (diagrams (d)-(f)). The diagrams with $D^+ \bar{D}^{\ast 0}$ as intermediate states are not presented, which are the same as those for $D^{\ast +} \bar{D}^0$. \label{fig:tri-zc}}
\end{figure}

With the above effective Lagrangians, we can calculate the hidden charm decay processes $Z_c \to J/\psi \pi$ and $Z_c \to \eta_c \rho$. As for $Z_c \to J/\psi \pi$, the amplitudes corresponding to diagrams Fig. \ref{fig:tri-zc}-(a)-(c) are
\begin{eqnarray}
\mathcal{M}_1&=&(i)^3\int \frac{d^4q}{(2\pi)^4}
    \big[g_{Z_c}\epsilon^{\phi}_{Z_c}\big]\big[-{ig_{D^*DP}}(ip_3^{\mu})\big]\nonumber\\
&&\times\big[-ig_{\psi DD}\epsilon_{\psi}^{\nu}
    (-iq_{\nu}+ip_{2\nu})\big]\frac{-g_{\mu\phi}+p_{1\mu}p_{1\phi}/m_{D^\ast}^2}{p_1^2-m_{D^\ast}^2}\nonumber\\
&& \times \frac{1}{p_2^2-m_{\bar D}^2}\frac{1}{q^2-m_D^2}\mathcal{F}(m_D,q^2),\nonumber\\
\mathcal{M}_2&=&(i)^3\int \frac{d^4q}{(2\pi)^4}
    \big[g_{Z_c}\epsilon^{\phi}_{Z_c}\big]\big[\frac12g_{D^*D^*P}\epsilon_{\mu\nu\alpha\beta}(ip_3^{\nu})\nonumber\\
&&\times(-ip_1^{\alpha}-iq^{\alpha})\big]\big[g_{\psi D^*D}\epsilon^{\tau}_{\psi}
    \epsilon_{\eta\tau\rho\sigma}(
    ip_4^{\eta})(-ip_2^\sigma+iq^{\sigma})\big]\nonumber\\
&& \times
    \frac{-g^{\phi\beta}+p_1^{\phi}p_1^{\beta}/m_{D^\ast}^2}{p_1^2-m_{D^\ast}^2}
    \frac{1}{p_2^2-m_{\bar D}^2}
    \frac{-g^{\mu\rho}+q^{\mu}q^{\rho}/m_{D^\ast}^2}{q^2-m_{D^\ast}^2}\nonumber\\
&&{\times}   \mathcal{F}(m_{D^\ast},q^2),\nonumber\\
\mathcal{M}_3&=&(i)^3\int \frac{d^4q}{(2\pi)^4}
   \big[g_{Z_c}\epsilon_{Z_c}^{\phi}\big]\big[-\frac{ig_{D^*DP}}{\sqrt{2}}(ip_3^{\mu})\big]\nonumber\\
&&\times \big\{ig_{\psi D^*D^*}
   \epsilon_{\psi}^{\rho}\big[g^{\tau}_\rho(-iq^\eta+ip_2^\eta)
-g^{\eta}_\rho(-ip_2^\tau+iq^\tau)\nonumber\\
&&    -g^{\tau\eta}(-iq_\rho+ip_{2\rho)}\big]\big\}\frac{1}{p_1^2-m_{\bar D}^2}\frac{-g_{\phi\tau}+p_{2\phi}p_{2\tau}/m_{D^\ast}^2}{p_2^2-m_{D^\ast}^2}\nonumber\\
&&\frac{-g_{\eta\mu}+q_{\eta}q_{\mu}/m_{D^\ast}^2}{q^2-m_{D^\ast}^2}\mathcal{F}(m_{D^\ast},q^2),
\end{eqnarray}
where a form factor $\mathcal{F}$ is introduced to reflect the off-shell effect and to make the amplitude convergent in the ultraviolet region.  The estimates in Ref.~\cite{Xiao:2017uve} indicate that the results are weakly dependent on the explicit form of the form factors. In the present work, we use the form factor of
\begin{eqnarray}
  \mathcal{F}(m,q^2)=\left(\frac{m^2-\Lambda^2}{q^2-\Lambda^2}\right)^3,
\end{eqnarray}
where the $\Lambda$ is reparameterized as $\Lambda=m_E+\alpha\Lambda_{\rm QCD}$, $m_E$ is the mass of the exchanged meson, and $\Lambda_{\rm QCD}=0.220$\,GeV. {The details about the form factor will be discussed in the Sect.~\ref{sec:3}.}

The total amplitude of $Z_c\to J/\psi\pi$ is
\begin{eqnarray}
  \mathcal M_{Z_c\to J/\psi\pi}^{\rm tot}=2 (\mathcal M_1+\mathcal M_2+\mathcal M_3),\label{eq:amp-zc-jpsipi}
\end{eqnarray}
where the factor 2 comes from the processes in which $D^{+} \bar{D}^{\ast 0}$ is the intermediate states. The partial width of the process $Z_c\to J/\psi\pi$ reads,
\begin{equation}
\Gamma(Z_c\to J/\psi\pi)=\frac{1}{3}\frac{1}{8\pi}\frac{|\vec p|}
    {m_{Z_c}^2}|\overline{{{\mathcal{M}}_{Z_c\to J/\psi\pi}^{\text{tot}}}}|^2, \label{Eq:PW}
\end{equation}
where the overline is the sum over the polarization of  $J/\psi$.

As for the process $Z_c\to \eta_c\rho$, the amplitudes corresponding to Fig. \ref{fig:tri-zc}-(d)-(f) read
\begin{eqnarray}
  \mathcal{M}_4&=&(i)^3\int \frac{d^4q}{(2\pi)^4}
    \big[g_{Z_c^\prime}\epsilon^{\phi}_{Z_c}\big]\big[ig_{D^*D^*V}\epsilon_\rho^\tau g^{\rho\eta}(-ip_1^\tau-iq^\tau)  \nonumber\\
&&\phantom{\times}+4if_{D^*D^*V}\epsilon_\rho^\tau
    (-ip_3^\eta g^{\tau\rho}+ip_3^\rho g^{\tau\eta})\big]\nonumber\\
&&\times\big[-ig_{\eta_c D^\ast D}(-ip_{2\mu}+iq_{\mu})\big]
    \frac{-g_{\eta\phi}+p_{1\eta}p_{1\phi}/m_{D^\ast}^2}{p_1^2-m_{D^\ast}^2}\nonumber\\
&&\times
\frac{1}{p_2^2-m_{\bar D}^2}
\frac{-g_{\mu\rho}+q_{\mu}q_{\rho}/m_{D^\ast}^2}{q^2-m_{D^\ast}^2}\mathcal{F}(m_{D^\ast},q^2),\nonumber\\
\mathcal{M}_5&=&(i)^3\int \frac{d^4q}{(2\pi)^4}
    \big[g_{Z_c^\prime}\epsilon^{\phi}_{Z_c}\big]\big[-ig_{DDV}(iq_\tau+ip_{1\tau})
    \epsilon_{\rho}^\tau\big]\nonumber\\
&&\times\big[-ig_{\eta_c D^*D}(-iq^\nu+ip_2^\nu)]\frac{1}{p_1^2-m_D^2}\nonumber\\
&&\times    \frac{-g_{\nu\phi}+p_{2\nu}p_{2\phi}/m_{D^\ast}^2}{p_2^2-m_{D^\ast}^2}
    \frac{1}{q^2-m_D^2}
    \mathcal{F}(m_D,q^2),\nonumber\\
\mathcal{M}_6&=&(i)^3\int \frac{d^4q}{(2\pi)^4}
    \big[g_{Z_c^\prime}\epsilon^{\phi}_{Z_c}\big]
    \big[-2f_{D^\ast DV}\epsilon_{\mu\tau\alpha\beta}
    (ip_{3}^\mu) \epsilon_\rho^\tau\nonumber\\
&&\times
    (iq^\alpha+ip_{1}^\alpha)\big]\big[-g_{\eta_c D^\ast D^\ast}\epsilon^{\eta\nu\psi\sigma} (ip_{4\eta})\nonumber\\
&&\times
   (-iq_{\psi}+ip_{2\psi})\big]\frac{1}{p_1^2-m_{\bar D}^2}\frac{-g_{\nu\phi}+p_{2\nu}p_{2\phi}/m_{D^\ast}^2}{p_2^2-m_{D^\ast}^2}
    \nonumber\\
&&\times\frac{-g^{\beta}_\sigma+q_{\sigma}q^{\beta}/m_{\bar D^\ast}^2}
    {q^2-m_{\bar D^\ast}^2}\mathcal{F}(m_{\bar D^\ast},q^2),
\end{eqnarray}
and the total amplitude for process $Z_c\to\eta_c\rho$ is
\begin{eqnarray}
  \mathcal M_{Z_c\to \eta_c\rho}^{\rm tot}=2 (\mathcal M_4+\mathcal M_5+\mathcal M_6).
\end{eqnarray}

Different from $\pi$, $\rho$ meson has a large width. Thus, when we estimate the partial width of $Z_c \to \eta_c \rho$, the effect of the $\rho$ meson width should be included. The partial width of $Z_c \to \eta_c \rho$ reads
\begin{eqnarray}
\Gamma_{Z_c\to \eta_c \rho}&=&  \frac{1}{W_\rho}\int\limits^{(m_{Z_c}-m_{\eta_c})^2}_{(2m_{\pi})^2}
    ds f(s,m_{\rho},\Gamma_{\rho}) \nonumber\\
    && \times
    \frac{|\vec p|}{24\pi m^2_{Z_c}}  |\overline{\mathcal M^{\text{tot}}_{Z_c\to \eta_c\rho}(m_{\rho}\to \sqrt s)}|^2,\label{eq:wid-etacrho}
\end{eqnarray}
where $W_\rho=\int_{(2m_\pi)^2}^{(m_{Z_c}-m_{\eta_c})^2}  ds f(s, m_\rho,\Gamma_\rho)$. $f(s,m_{\rho},\Gamma_{\rho})$ is a relativistic form of the Breit-Wigner distribution, which reads
\begin{eqnarray}
f(s,m_{\rho},\Gamma_{\rho})=\frac{1}{\pi}\frac{m_{\rho}\Gamma_{\rho}}{(s-m_{\rho}^2)^2+m_{\rho}^2\Gamma_{\rho}^2},
\label{eq:breit-wigner}
\end{eqnarray}
where $m_\rho=775$\,MeV and $\Gamma_\rho=149$\,MeV are the mass and width of the $\rho$ meson {\cite{Tanabashi:2018oca}}.

\subsection{The hidden charm decays of $Z_c^{\prime}$}
\begin{figure*}[hbt!]
\begin{tabular}{cccc}
\includegraphics[scale=0.35]{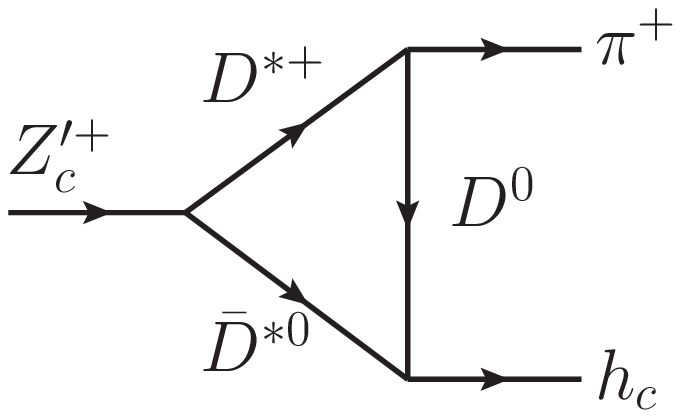}
&\includegraphics[scale=0.35]{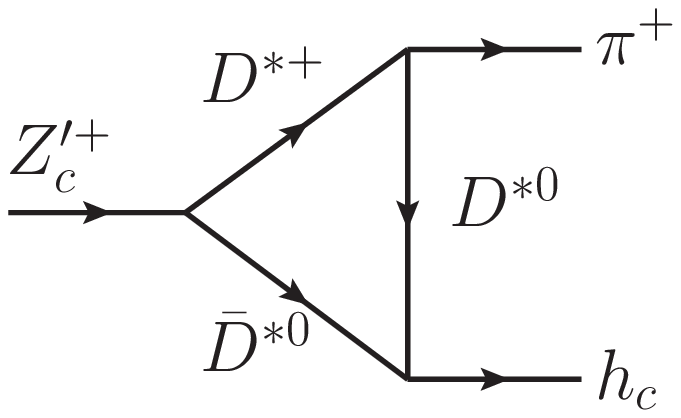}
&\includegraphics[scale=0.35]{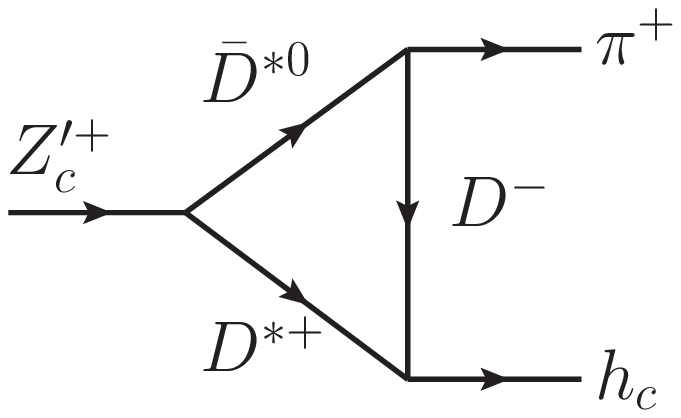}
&\includegraphics[scale=0.35]{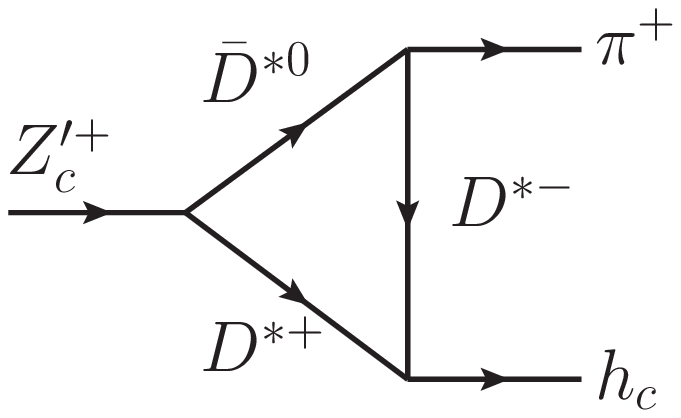}\\
(a) &(b) & (c)    &(d) \\ \\
\includegraphics[scale=0.35]{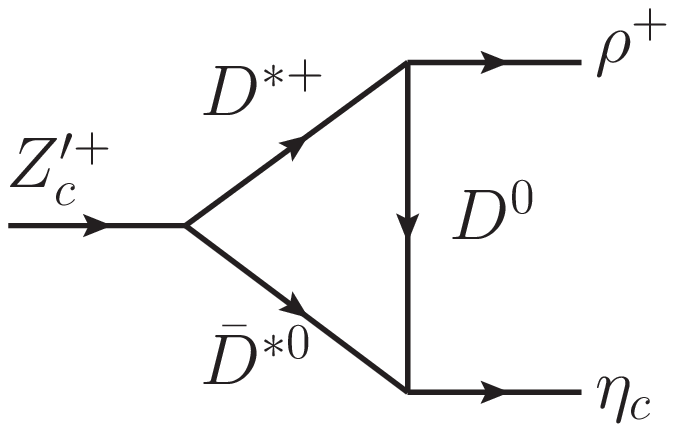}
&\includegraphics[scale=0.35]{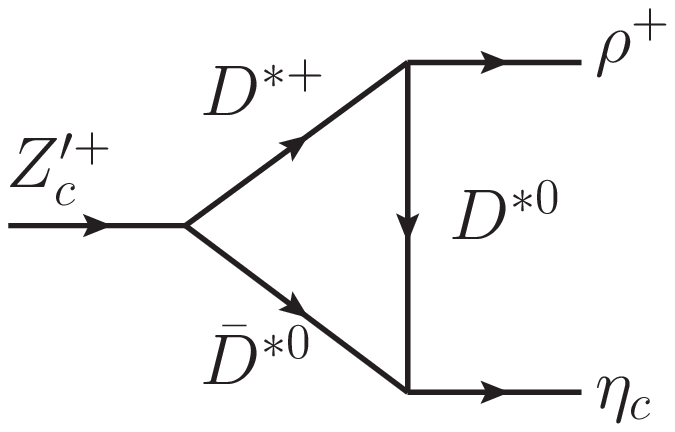}
&\includegraphics[scale=0.35]{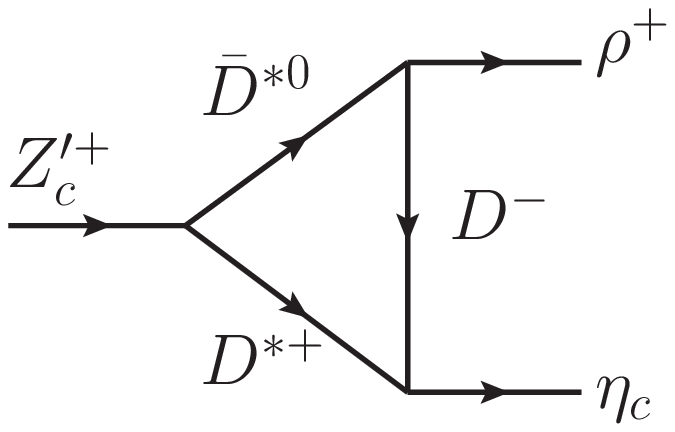}
&\includegraphics[scale=0.35]{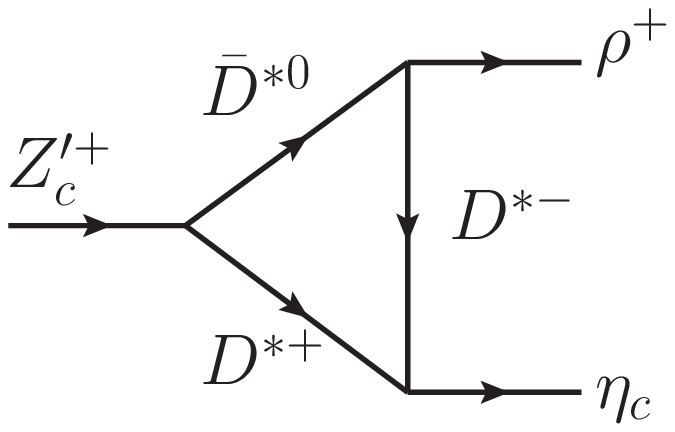}\\
(e) &(f) &
(g)    &(h)
  \end{tabular}
\caption{Diagrams contributing to the processes $Z_c^{\prime}\to h_c\pi^+$ and $Z_c^\prime \to \eta_c \rho^+$.\label{fig:tri-zcp}}
\end{figure*}

The $Z_c^\prime$ is first observed in the $h_c \pi$ channel and the new measurement from BESIII also reported the  ratio of partial widths between $\eta_c \rho$ and $h_c \pi$ modes. Thus, we estimate these two decay modes of $Z_c^\prime$ with triangle loop mechanism in the present work and the corresponding sketch diagrams of these two channels are shown in Fig. \ref{fig:tri-zcp}. The amplitudes of $Z_c^\prime \to h_c \pi$ corresponding to Fig. \ref{fig:tri-zcp}-(a)-(b) read
\begin{eqnarray}
  \mathcal{M}_a&=&(i)^3\int \frac{d^4q}{(2\pi)^4}
    \big[ig_{Z_c^\prime}
   \epsilon_{\mu\nu\alpha\beta}(-ip_0^\mu)
   \epsilon_{Z_c^\prime}^{\nu}\big]\big[-ig_{D^\ast DP}(ip_{3\phi})\big]\nonumber\\
&&\times\big[g_{h_c D^\ast D}\epsilon_{h_c}^\psi\big]
    \frac{-g^{\alpha\phi}+p_1^{\alpha}p_1^{\phi}/m_{D^\ast}^2}
    {p_1^2-m_{D^\ast}^2}\nonumber\\
&&\times
    \frac{-g^{\beta}_\psi+p_2^{\beta}p_{2\psi}/m_{\bar D^\ast}^2}{p_2^2-m_{\bar D^\ast}^2}
    \frac{1}{q^2-m_{D}^2}\mathcal{F}(m_D,q^2),\nonumber\\
\mathcal{M}_b&=&(i)^3\int \frac{d^4q}{(2\pi)^4}
    \big[ig_{Z_c^\prime}
   \epsilon_{\mu\nu\alpha\beta}(-ip_0^\mu)\epsilon_{Z_c^\prime}^{\nu}\big]
   \big[\frac12g_{D^\ast D^\ast P}\epsilon_{\eta\tau\rho\sigma}\nonumber\\
&&\times(ip_3^\tau)(-ip_1^\rho-iq^\rho)\big]
    \big[ig_{h_c D^\ast D^\ast}\epsilon_{\psi\phi\delta\gamma}
    (ip_4^\psi)\epsilon_{h_c}^\phi\big]\nonumber\\
&&\times\frac{-g^{\sigma\alpha}+p_1^{\sigma}p_1^{\alpha}/m_{D^\ast}^2}{p_1^2-m_{D^\ast}^2}
    \frac{-g^{\beta\gamma}+p_2^{\beta}p_2^{\gamma}/m_{\bar D^\ast}^2}{p_2^2-m_{\bar D^\ast}^2}\nonumber\\
&&\times    \frac{-g^{\delta\eta}+q^{\delta}q^{\eta}/m_{D^\ast}^2}
    {q^2-m_{D^\ast}^2} \mathcal{F}(m_{D^\ast},q^2).
\end{eqnarray}
The amplitudes $\mathcal{M}_{c}$ and $\mathcal{M}_d$ corresponding to Fig. \ref{fig:tri-zcp}-(c)-(d) are the same as $\mathcal{M}_a$ and $\mathcal{M}_b$, respectively. Therefore, the total amplitude for the process $Z_c^\prime\to h_c\pi$ is
\begin{eqnarray}
  \mathcal M_{Z_c^\prime\to h_c\pi}^{\rm tot}=2(\mathcal M_a+\mathcal M_b)\label{eq:amp-zcp-hcpi}.
\end{eqnarray}

The amplitudes of $Z_c^\prime \to \eta_c \rho$ corresponding to Fig. \ref{fig:tri-zcp}-(e)-(f) are
\begin{eqnarray}
  \mathcal{M}_e&=&(i)^3\int \frac{d^4q}{(2\pi)^4}
    \big[ig_{Z_c^\prime}
   \epsilon_{\mu\nu\alpha\beta}(-ip_0^\mu)\epsilon_{Z_c^\prime}^{\nu}\big]
  \big[ -2f_{D^\ast DV}\epsilon_{\eta\tau\rho\sigma}\nonumber\\
 &&\times   (ip_3^\eta)\epsilon_{\rho}^\tau(-ip_1^\rho-iq^\rho)\big]
    \big[-ig_{\eta_c D^\ast D}(-ip_2^\phi+iq^\phi)\big]\nonumber\\
&&\times\frac{-g^{\sigma\alpha}+p_1^{\sigma}p_1^{\alpha}/m_{D^\ast}^2}
    {p_1^2-m_{D^\ast}^2}
    \frac{-g^{\beta}_{\phi}+p_{2\phi}p_2^{\beta}/m_{\bar D^\ast}^2}{p_2^2-m_{\bar D^\ast}^2}\nonumber\\
&&\times
    \frac{1}{q^2-m_{D}^2}
    \mathcal{F}(m_D,q^2),\nonumber
\end{eqnarray}
\begin{eqnarray}
\mathcal{M}_f&=&(i)^3\int \frac{d^4q}{(2\pi)^4}
    \big[ig_{Z_c^\prime}
   \epsilon_{\mu\nu\alpha\beta}(-ip_0^\mu)\epsilon_{Z_c^\prime}^{\nu}\big]
   ig_{D^\ast D^\ast V}\epsilon_\rho^\tau g^{\sigma\eta}\nonumber\\
&&\times(-ip_1^\tau-iq^\tau)
    +4if_{D^\ast D^\ast V}\epsilon_\rho^\tau(-ip_3^\eta g^{\tau\sigma}
    +ip_3^\sigma g^{\tau\eta})\big]\nonumber\\
&&\phantom{\times}
    \big[-g_{\eta_c D^\ast D^\ast}\epsilon_{\lambda\delta\psi\phi}
    (-iq^\lambda)(ip_4^\phi)\big]\nonumber\\
&&\times\frac{-g^{\eta\alpha}+p_1^{\eta}p_1^{\alpha}/m_{D^\ast}^2}{p_1^2-m_{D^\ast}^2}
    \frac{-g^{\beta\psi}+p_2^{\beta}p_2^{\psi}/m_{\bar D^\ast}^2}{p_2^2-m_{\bar D^\ast}^2}\nonumber\\
&&\times    \frac{-g^{\delta\sigma}+q^{\delta}q^{\sigma}/m_{D^\ast}^2}
    {q^2-m_{D^\ast}^2}\mathcal{F}(m_{D^\ast},q^2). \label{eq:amp-zcp-etacrho}
\end{eqnarray}
Similar to the case of $Z_c^\prime \to h_c \pi$ , the amplitudes $\mathcal{M}_{g}$ and $\mathcal{M}_{h}$ corresponding to Fig. \ref{fig:tri-zcp}-(g)-(h) are the same as $\mathcal{M}_{e}$ and $\mathcal{M}_f$, respectively. The total amplitude for the process $Z_c^\prime\to \eta_c\rho$ is
\begin{eqnarray}
  \mathcal M_{Z_c^\prime\to h_c\pi}^{\rm tot}=2(\mathcal M_e+\mathcal M_f)\label{eq:amp-zcp-hcpi}.
\end{eqnarray}
With the help of Eqs. (\ref{Eq:PW}) and (\ref{eq:wid-etacrho}), one can estimate the partial widths of $Z_c^\prime \to J/\psi \pi$ and $Z_c^\prime \to \eta_c \rho$.

\section{Numerical Results and Discussion}\label{sec:3}

\begin{figure}[hbt!]
  \includegraphics[scale=0.9]{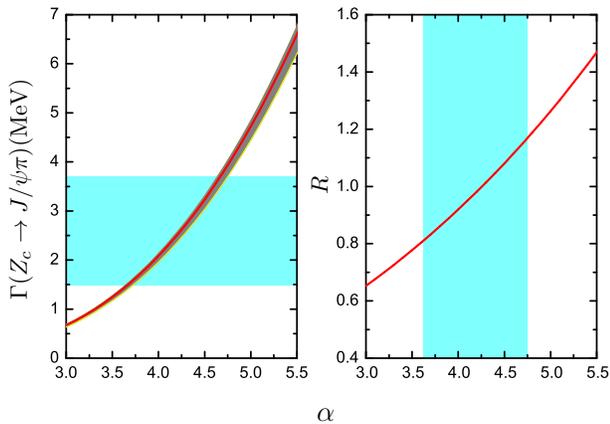}
  \caption{The partial width for the process $Z_c\to J/\psi\pi$ (left panel) and the ratio $R$ of Eq. (\ref{eq:ratio-etacrho-jpsipi}) (right panel) depend on the model parameter $\alpha$. The red curve with gray band in the left is the estimated result in the present work and the gray band indicates the errors resulted from the uncertainty of $g_{Z_c D^\ast D}$. The cyan horizontal band in the left is the partial width extracted from the experimental data. The vertical band in the right is the $\alpha$ range determined by the partial width of $Z_c\to J/\psi \pi$.\label{fig:zc}}
\end{figure}

\renewcommand\arraystretch{1.5}
\begin{table*}[hbt!]
\caption{The relevant masses and coupling constants in the present calculations. In the heavy quark limit , the coupling constants $g_{\psi D^{(\ast)} D^{(\ast)}}$  can be related to the gauge coupling $g_2$ and is determined to be $\sqrt{m_\psi}/(2m_Df_\psi)$ by the VMD model in the process $e^- D^+\to e^-D^+$\cite{Lin:1999ad,Oh:2000qr}. Here, $f_\psi=416$\,MeV is the $J/\psi$ decay constant, which is determined by the experimental partial width $\Gamma(J/\psi\to e^+e^-)=5.55\,$keV\cite{Tanabashi:2018oca}. As for the $P$-wave charmonium and charmed meson interactions, the coupling constants can be related to the gauge coupling $g_1$, where $g_1=-\sqrt{m_{\chi_{c0}}/3}/f_{\chi_{c0}}$ with $f_{\chi_{c0}}=0.51$ \,GeV~\cite{Colangelo:2003sa}. In the couplings of charmed meson and light meson, the gauge coupling $g=0.55$ is determined via the partial decay width $\Gamma(D^\ast\to D\pi)$ \cite{Tanabashi:2018oca}, and $f_\pi=132$\,MeV is the pion decay constant. The parameters of couplings related to the light vector mesons are $\beta=0.9$, $\lambda=0.56\, \rm{GeV}^{-1}$, and $g_V=m_\rho/f_\pi$\cite{Isola:2003fh,Chen:2010re}.  \label{tab:coupling}}
  \begin{tabular}{cccccccc}
    \toprule[1pt]
    Meson & $Z_c^{(\prime)}$  & $J/\psi$ &$\eta_c$  & $h_c$ & $D^{\ast+(0)}$ & $D^{+(0)}$ & $\pi^+$ \\
    Mass(GeV) & 3.887(4.024) & 3.097 &2.984  & 3.525 & 2.010(2.007) & 1.869(1.864) & 0.140 \\
   \midrule[1pt]
Couplings & $g_{\psi DD}$ & $ g_{\psi D^\ast D}$ & $g_{\psi D^\ast D^\ast}$ & $g_{\eta_c D^\ast D}$ & $g_{\eta_c D^\ast D^\ast}$  &$g_{h_c D^\ast D}$ & $g_{h_c D^\ast D^\ast}$\\
Expressions &$2g_2\sqrt{m_{J/\psi}}m_D$   &$2g_2\sqrt{m_Dm_{D^\ast}/m_\psi}   $ &$2g_2\sqrt{m_\psi}m_{D^\ast}  $  &$2g_2\sqrt{m_{\eta_c}m_{D^\ast}m_D} $ &$2g_2{m_{D^\ast}}/\sqrt{m_{\eta_c}}$   &$2g_1\sqrt{m_{h_c}m_Dm_{D^\ast}}$  &$2g_1m_{D^\ast}/\sqrt{m_{h_c}}$\\
Values &7.41   &3.98\,GeV$^{-1}$   &7.98   &7.55   &2.63\,GeV$^{-1}$   &-15.19\,GeV &-4.47\\
\midrule[1pt]
Couplings &  $g_{DDV}$ & $g_{D^\ast DV}$ &$g_{D^\ast D^\ast V}$   &$f_{D^\ast D^\ast V}$  & $g_{D^\ast DP}$ &$g_{D^\ast D^\ast P}$\\
Expressions  & $\beta g_V/\sqrt2$ & $\lambda g_V/\sqrt2$ &$\beta g_V/\sqrt2$ &$\lambda m_{D^\ast}g_V\sqrt2$  &${2g}\sqrt{m_{D^\ast}m_D}/{f_\pi}$ &$ {2g}/{f_\pi}$\\
Values &3.71    &2.30\,GeV$^{-1}$   &3.71   &4.64   &16.1   &8.33\,GeV$^{-1}$\\
\bottomrule[1pt]
  \end{tabular}
\end{table*}
{
To estimate the partial widths of considered processes,  the relevant coupling constants should be fixed.  The couplings between the $Z_c^{(\prime)}$ and charmed mesons can be determined by the experimental measured partial decay width of corresponding open charm modes, which can be obtained by total decay width and branching fractions of corresponding open charm modes. However, the experimental measured total decays are quite inaccurate
\begin{eqnarray}
  \Gamma^{\text{tot}}_{Z_c}&=&28.2\pm2.6\,\text{MeV},\label{eq:wid-tot-zc}\\
  \Gamma^{\text{tot}}_{Z_c^\prime}&=&13\pm5\,\text{MeV}.\label{eq:wid-tot-zcp}
\end{eqnarray}
Here, we apply the center values and the effect of the errors will be discussed in the later sentences. In addition, we assume that $Z_c$ dominantly decays into $(D^\ast \bar D+\text{H.c.})$ and $J/\psi\pi$. With the experimental measured ratio $R$ between the widths of $Z_c \to D^\ast \bar{D}$ and $Z_c \to J/\psi \pi$ {\cite{Ablikim:2013xfr}}
\begin{eqnarray}
 \frac{\Gamma(Z_c\to \bar D^\ast D)}{\Gamma ( Z_c \to J/\psi\pi)}=6.2\pm1.1\pm2.7.\label{eq:ratio-dd-jpsipi}
\end{eqnarray}
we can approximately obtain the branching fractions of the open charm modes. Finally, the coupling constant $g_{Z_c}$ is determined to be $1.13$ GeV with the center values of the ratio in Eq.~(\ref{eq:ratio-dd-jpsipi}), while the effect of the errors of the ratio in Eq.~(\ref{eq:ratio-dd-jpsipi}) will be discussed in the later sentences.
}

As for the $Z_c^\prime$, the Born cross sections for $e^+ e^-\to \pi^\pm (D^\ast\bar{D}^\ast)^\mp$ at 4.26 GeV are measured to be $137 \pm 9 \pm15$ pb and the fraction from the quasi-two-body cascade decay is {\cite{Ablikim:2013emm}}
\begin{eqnarray}
\label{eq:ratio-sigma-4020}
&&\frac{\sigma(e^+e^-\to\pi^\pm Z_c(4020)^{\mp}\to\pi^\pm (D^\ast\bar D^\ast)^\mp)}
  {\sigma(e^+e^-\to\pi^\pm (D^\ast\bar D^\ast)^\mp)}\nonumber \\
&&=0.65\pm0.09\pm0.06.
\end{eqnarray}
At the same energy point, the cross section of the quasi-two-body process $e^+e^-\to\pi^\pm Z_c(4020)^{\mp}\to\pi^+\pi^-h_c$ is measured to be  {\cite{Ablikim:2013wzq}}
\begin{eqnarray}
\sigma(e^+e^-\to\pi^\pm Z_c(4020)^{\mp}\to\pi^+\pi^-h_c)=7.4\pm1.7\pm2.1\ \rm{pb}. \nonumber\\
\end{eqnarray}
Therefore, the ratio between widths of $Z_c^\prime \to D^\ast \bar{D}^\ast$ and $Z_c^\prime \to h_c \pi$ is estimated
\begin{eqnarray}
\frac{\Gamma(Z_c^\prime \to D^\ast \bar{D}^\ast)}{\Gamma(Z_c \to h_c \pi)} =12.0\pm3.68\pm3.48.\label{eq:ratio-ddstar-hcpi}
\end{eqnarray}
{Similar to the $g_{Z_c}$,  one can get the coupling constant $g_{Z_c^\prime}= 5.63$ with the center values of the total decay width of $Z_c^\prime$ in Eq.~(\ref{eq:wid-tot-zcp}) and the ratio in Eq.~(\ref{eq:ratio-ddstar-hcpi}).}

The coupling constants of  charmonia (light mesons) and charmed mesons can be related to some gauge coupling constants in the heavy quark limit and chiral symmetry, which are listed in Table. \ref{tab:coupling}. {Besides the coupling constants, in the present calculation there is one more parameter, $\alpha$, which is introduced by the form factor.  The form factor is adopted to represent the off-shell effect of the exchanging particle.  In Ref.~\cite{Gortchakov:1995im}, a monopole-type form factor, $(m^2-\Lambda^2)/(q^2-\Lambda^2)$, was preferred based on the QCD sum rule calculation. However, the monopole form is not the unique one. In Ref.~\cite{Cheng:2004ru}, the authors applied two types of form factor, the monopole and dipole, which is the square of the monopole and  they consider the values of $\alpha$ should be in order of a magnitude of 1. In our previous work\cite{Xiao:2017uve}, we considered some different kinds of form factors and  found that a similar result can be obtained with different forms of form factor, while the $\alpha$ varies in different form factors. Actually, the value of $\alpha$ can not be determined from the first principle. Alternatively, it can be fixed via experimental data.
}

With the above preparation, we can estimate the partial widths of the hidden charm decays of $Z_c^{(\prime)}$. In Fig.~\ref{fig:zc},  the partial width for the process $Z_c\to J/\psi\pi$  and the ratio $R$ depending on parameter $\alpha$ are presented. The estimated error of $\Gamma(Z_c \to J/\psi \pi)$ results from the uncertainty of $g_{Z_c D^\ast D}$, which is determined by the partial width of $Z_c \to D^\ast \bar{D}$. {As mentioned above, the partial width of $Z_c\to D^\ast \bar{D}$ is estimated by the measured total decay width of $Z_c$ and the ratio in Eq.~(\ref{eq:ratio-dd-jpsipi}). In the estimation  of the coupling constants, we only consider the error in Eq. (\ref{eq:ratio-dd-jpsipi}), while the error of the total width in Eq.~(\ref{eq:wid-tot-zc}) is not included since this error will not affect the ratio $R$.  The cyan horizontal band is the experimental partial width of the $Z_c \to J/\psi \pi$ process, which is obtained by the center value of the total width and the ratio in Eq. (\ref{eq:ratio-dd-jpsipi}) with the assumption that $Z_c$ dominantly decays into $D^\ast \bar{D}+c.c$ and $J/\psi \pi$. In addition, this experimental information can be used to determine the value of parameter $\alpha$ in the hidden charm decays of $Z_c$.
}

By comparing our estimation with the experimental data, one can find in the range of $\alpha=3.63 \sim 4.75$, that the experimental data can be reproduced.
In the same way, we can estimate the partial width of $Z_c \to \eta_c \rho$ and the ratio of the widths of $Z_c\to \eta_c \rho$ and $Z_c \to J/\psi \pi$. In the right panel of Fig.~\ref{fig:zc}, we present the $\alpha$ dependence of $R$. Within the $\alpha$ range determined by the partial width of $Z_c \to J/\psi \pi$, the ratio is determined to be 0.81$-$1.17.  This ratio is very close to the lower limit of the preliminary results from the BESIII Collaboration, which is $2.1 \pm 0.8$. {Our results indicate that the triangle loop mechanism plays a dominant role in understanding the hidden charm decays of $Z_c$.
}

\begin{table}[hbt!]
\caption{The partial widths of $J/\psi\pi/\eta_c\rho$ modes and the ratio $R$ for $Z_c$ estimated in different frames \cite{Goerke:2016hxf,  Agaev:2016dev, Dias:2013xfa, Wang:2017lot, Esposito:2014hsa, Ke:2013gia}.\label{tab:width-zc}}
  \begin{tabular}{ccccc}
    \toprule[1pt]
Model  & Ref.  &$\Gamma(J/\psi\pi)$(MeV)    &$\Gamma(\eta_c\rho)$(MeV)   &$R$\\\hline
\multirow{6}{*}{Tetraquark}    &  \cite{Goerke:2016hxf}   &$27.9^{+6.3}_{-5.0}$   &$35.7^{+6.3}_{-5.2}$&$\cdots$\\
   &\cite{Agaev:2016dev}   &$41.9\pm9.4$   &$65.7\pm10.6$  &$\cdots$\\
   &\cite{Dias:2013xfa}     &$29.1\pm8.2$   &$27.5\pm8.5$   &$\cdots$\\
   &\cite{Wang:2017lot}   &$25.8\pm9.6$   &$27.9\pm20.1$   &$\cdots$\\
   &\cite{Esposito:2014hsa}   &$\cdots$   &$\cdots$   &$230^{+330}_{-140}$\\
   &\cite{Esposito:2014hsa}  &$\cdots$   &$\cdots$   &$0.27^{+0.40}_{-0.17}$  \\
    \midrule[1pt]
\multirow{3}{*}{Molecule}  &\cite{Goerke:2016hxf}  &$1.8\pm0.3$    &$3.2^{+0.5}_{-0.4}$    &$\cdots$\\
    &  \cite{Esposito:2014hsa}   &$\cdots$   &$\cdots$   &$0.046^{+0.025}_{-0.017}$\\
    & \cite{Ke:2013gia}  &$3.67$     &0.45   &$\cdots$\\
 \midrule[1pt]
  Present & & 1.47$-$3.71&1.19$-$4.34 & 0.81$-$1.17 \nonumber\\
\bottomrule[1pt]
  \end{tabular}
\end{table}

For comparison, we also present the partial widths of $Z_c \to \eta_c \rho/J/\psi \pi$ and their ratio estimated from different methods in Table~\ref{tab:width-zc}  \cite{Goerke:2016hxf,  Agaev:2016dev, Dias:2013xfa, Wang:2017lot, Esposito:2014hsa, Ke:2013gia}. {As we discussed at the end of the Introduction, the hidden charm decays of the tetraquark state should occur dominantly via the quark rearrangement, and the contributions from the quark rearrangement should be much larger than the ones from the triangle loop mechanism although the triangle loop mechanism always exists in both tetraquark and molecular scenarios. In the present work, the  partial widths of hidden charm decay modes  resulting from the triangle loop mechanism are estimated to be of order of several MeV, while in the tetraquark scenario, the hidden charm decay widths   were evaluated in the literatures to be several tens MeV  by using different methods \cite{Goerke:2016hxf,  Agaev:2016dev, Dias:2013xfa, Wang:2017lot}, which are at least 1 order of magnitude larger than the ones from the triangle loop mechanism. Such a conclusion is consistent with our analysis. Furthermore, our estimated partial widths of $\eta_c \rho/ J/\psi \pi$ and their ratio are comparable with the experimental data, which indicates that the triangle loop mechanism plays a dominant role in understanding the hidden bottom decays of $Z_c$, and thus the present estimation supports $Z_c$ as a molecular state.}

\begin{figure}[hbt!]
\includegraphics[scale=0.90]{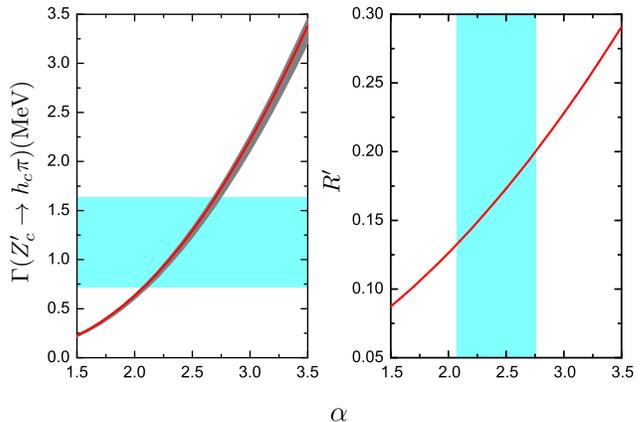}
\caption{The figures for the process $Z_c^\prime\to h_c\pi$ and the ratio $R^\prime$ of Eq.~(\ref{eq:ratio-etacrho-hcpi}) similar to Fig.~\ref{fig:zc}.\label{fig:zcprime}}
\end{figure}

As for $Z_c^\prime$, we take the first observed channel $Z_c^\prime \to h_c \pi$ as a scale to determine the $\alpha$ range. The partial width of $Z_c^\prime \to h_c \pi$ depending on $\alpha$ is presented in the left panel of Fig. \ref{fig:zcprime}. Using the same way to determine the partial widths of $Z_c$, we extract the partial width of $Z_c^\prime \to h_c \pi$  to be $(0.72\sim1.63)\  \mathrm{MeV} $ via the experimental data \cite{Ablikim:2013emm}. Our estimated result overlaps with this data in the $\alpha$ range of $2.07 \sim 2.75$. In this determined $\alpha$ range, the ratio of the partial widths of $Z_c^\prime \to \eta_c \rho$ and $Z_c^\prime \to h_c \pi$ is estimated to be
$0.13\sim0.20$, which is safely under the measured upper limit 1.9 \cite{Yuan:2018inv}. {In other words, the triangle loop mechanism is also found important in the decays of $Z_c^\prime$. Similar to the analysis of the hidden charm decays of $Z_c$, we conclude that $Z_c^\prime$ can be interpreted as the hadronic molecule, while the tetraquark interpretation is less favored.} Bedsides, there are also results from other works, where the author in Ref.~\cite{Esposito:2014hsa} estimated the $R^\prime$ to be $6.6^{+56.8}_{-5.8}$ and $0.010^{+0.006}_{-0.004}$ in the tetraquark and molecule models, respectively, which are quite different from our estimate in this paper.

\section{Summary}\label{sec:4}

{
 Stimulated by the recent measurements of $Z_c^{(\prime)}\to \eta_c \rho$ reported by the BESIII Collaboration, we estimated the hidden charm decay processes and tried to lift the curtain on the $Z_c$ and $Z_c^\prime$. We have noticed that both $Z_c$ and $Z_c^\prime$ dominantly decay into a pair of charmed
mesons, and the charmed meson pair can transit into a charmonium and a light meson by exchanging a proper charmed
meson. Such a triangle loop mechanism can be used to distinguish the nature of the $Z_c$ and $Z_c^\prime$.}

As for $Z_c$, we have considered the discovered channel $Z_c \to J/\psi \pi$ as a scale to determine the $\alpha$. In the range $\alpha=3.63 \sim 4.75$, our estimation is in agreement with the experimental measured partial width of the $Z_c \to J/\psi \pi$ process. This experimental decay width is approximately obtained since the experimental branching fractions are  absent. With this determined $\alpha $, we have found that the ratio of the partial widths of $Z_c \to \eta_c \rho$ and $Z_c\to J/\psi \pi$ is $0.81 \sim 1.17$, which is very close to the measured one. As for $Z_c^\prime$, taking the discovered channel $Z_c^\prime \to h_c \pi $ as a scale, we have found that our estimated partial width of $Z_c^\prime \to h_c \pi$ can overlap with the experimental data in the range $\alpha=2.07 \sim 2.75$. In this $\alpha$ range, the ratio of the partial widths of $Z_c^\prime \to \eta_c \rho$ and $Z_c \to h_c \pi$ is estimated to be $0.13\sim 0.20$, which is safely under the upper limit of the measurement from the BESIII Collaboration.

{Our estimations indicate that the triangle loop mechanism plays an important role in understanding the hidden charm decays of $Z_c$ and $Z_c^\prime$. Such decay behaviors of $Z_c$ and $Z_c^\prime$ are consistent with the hadronic molecule interpretation for these two $Z_c$ states; thus, the $Z_c$ and $Z_c^\prime$ can be assigned as the hadronic molecules. We also pointed out that the tetraquark interpretation is less favored, which is supported by the estimation of the hidden charm partial width in the QCD sum rule \cite{Goerke:2016hxf,  Agaev:2016dev, Dias:2013xfa, Wang:2017lot}.}

\section*{Acknowledgement}

This work is supported in part by the National Natural Science Foundation of China (NSFC) under Grants No.~11775050, No.~11375240, No.~11475192 and No.~11435014, by the fund provided to the Sino-German CRC 110 ``Symmetries and the Emergence of Structure in QCD" project by the NSFC under Grant No.11621131001, by the Key Research Program of Frontier Sciences, CAS, Grant No. Y7292610K1, and by the Fundamental Research Funds for the Central Universities.



\begin{thebibliography}{99}


\bibitem{Ablikim:2013mio}
  M.~Ablikim {\it et al.} (BESIII Collaboration),
  Observation of a Charged Charmoniumlike Structure in $e^+e^-\to$  $\pi^+\pi^-J/\psi$  at $\sqrt{s}$ =4.26\,GeV,
  Phys.\ Rev.\ Lett.\  {\bf 110}, 252001 (2013).

\bibitem{Liu:2013dau}
  Z.~Q.~Liu {\it et al.} (Belle Collaboration),
  Study of $e^+e^- \to \pi^+ \pi^- J/\psi$ and Observation of a Charged Charmoniumlike State at Belle,
  Phys.\ Rev.\ Lett.\  {\bf 110}, 252002 (2013).

\bibitem{Xiao:2013iha}
  T.~Xiao, S.~Dobbs, A.~Tomaradze and K.~K.~Seth,
  Observation of the charged hadron $Z_c^{\pm}(3900)$ and evidence for the neutral $Z_c^0(3900)$ in $e^+e^-\to \pi\pi J/\psi$ at $\sqrt{s}=4170$ MeV,
  Phys.\ Lett.\ B {\bf 727}, 366 (2013).

\bibitem{Ablikim:2013wzq}
  M.~Ablikim {\it et al.} (BESIII Collaboration],
  Observation of a Charged Charmoniumlike Structure $Z_c$(4020) and Search for the $Z_c$(3900) in $e^+e^- \to \pi^+\pi^-h_c$,
  Phys.\ Rev.\ Lett.\  {\bf 111}, 242001 (2013).


\bibitem{Ablikim:2014dxl}
  M.~Ablikim {\it et al.} (BESIII Collaboration),
  Observation of $e^+e^- \to\pi^0\pi^0h_c$ and a Neutral Charmoniumlike Structure $Z_c(4020)^0$,
  Phys.\ Rev.\ Lett.\  {\bf 113}, 212002 (2014).

\bibitem{Ablikim:2013xfr}
  M.~Ablikim {\it et al.} (BESIII Collaboration),
  Observation of a Charged $(D\bar{D}^{*})^\pm$ Mass Peak in $e^{+}e^{-} \to \pi D\bar{D}^{*}$ at $\sqrt{s} =$ 4.26 GeV,
  Phys.\ Rev.\ Lett.\  {\bf 112}, 022001 (2014).

\bibitem{Collaboration:2017njt}
  M.~Ablikim {\it et al.} (BESIII Collaboration),
  Determination of the Spin and Parity of the $Z_c(3900)$,
  Phys.\ Rev.\ Lett.\  {\bf 119}, 072001 (2017).

\bibitem{Ablikim:2013emm}
  M.~Ablikim {\it et al.} (BESIII Collaboration),
 Observation of a Charged Charmoniumlike Structure in $e^+e^- \to (D^{*} \bar{D}^{*})^{\pm} \pi^\mp$ at $\sqrt{s}=4.26$GeV,
  Phys.\ Rev.\ Lett.\  {\bf 112}, 132001 (2014).






\bibitem{Chen:2010ze}
  W.~Chen and S. L.~Zhu,
The vector and axial-vector charmonium-like States,
  Phys.\ Rev.\ D {\bf 83}, 034010 (2011).


\bibitem{Faccini:2013lda}
  L.~Maiani, V.~Riquer, R.~Faccini, F.~Piccinini, A.~Pilloni and A.~D.~Polosa,
A $J^{PG}=1^{++}$ charged resonance in the $Y(4260) \to \pi^+ \pi^- J/\psi$ decay?,
  Phys.\ Rev.\ D {\bf 87}, 111102 (2013).

\bibitem{Braaten:2013boa}
  E.~Braaten,
  How the $Z_c(3900)$ Reveals the Spectra of Quarkonium Hybrid and Tetraquark Mesons,
  Phys.\ Rev.\ Lett.\  {\bf 111}, 162003 (2013).


\bibitem{Goerke:2016hxf}
  F.~Goerke, T.~Gutsche, M.~A.~Ivanov, J.~G.~Korner, V.~E.~Lyubovitskij and P.~Santorelli,
Four-quark structure of $Z_c(3900)$, $Z(4430)$ and $X_b(5568)$ states,
  Phys.\ Rev.\ D {\bf 94}, 094017 (2016).



\bibitem{Qiao:2013raa}
  C.~F.~Qiao and L.~Tang,
  Estimating the mass of the hidden charm $1^+(1^{+})$ tetraquark state via QCD sum rules,
  Eur.\ Phys.\ J.\ C {\bf 74}, 3122 (2014).



\bibitem{Wang:2013vex}
  Z.~G.~Wang and T.~Huang,
  Analysis of the $X(3872)$, $Z_c(3900)$ and $Z_c(3885)$ as axial-vector tetraquark states with QCD sum rules,
  Phys.\ Rev.\ D {\bf 89}, 054019 (2014).




\bibitem{Deng:2014gqa}
  C.~Deng, J.~Ping and F.~Wang, Interpreting $Z_c(3900)$ and $Z_c(4025)/Z_c(4020)$ as charged tetraquark states,
  Phys.\ Rev.\ D {\bf 90}, 054009 (2014).

\bibitem{Agaev:2016dev}
  S.~S.~Agaev, K.~Azizi and H.~Sundu,
Strong $Z_c^{+}(3900)\rightarrow J/\psi \pi^{+}; \eta_{c} \rho^{+}$ decays in QCD,
  Phys.\ Rev.\ D {\bf 93}, 074002 (2016).



\bibitem{Dias:2013xfa}
  J.~M.~Dias, F.~S.~Navarra, M.~Nielsen and C.~M.~Zanetti,
  $Z^+_c$(3900) decay width in QCD sum rules,
  Phys.\ Rev.\ D {\bf 88}, 016004 (2013).

\bibitem{Wang:2017lot}
  Z.~G.~Wang and J.~X.~Zhang, The decay width of the $Z_c(3900)$ as an axialvector tetraquark state in solid quark-hadron duality,
  Eur.\ Phys.\ J.\ C {\bf 78}, 14 (2018).




\bibitem{Esposito:2014hsa}
  A.~Esposito, A.~L.~Guerrieri and A.~Pilloni,
Probing the nature of $Z_c^{(\prime)}$ states via the $\eta_c\rho$ decay,
  Phys.\ Lett.\ B {\bf 746}, 194 (2015).



\bibitem{Qiao:2013dda}
  C.~F.~Qiao and L.~Tang,
  Interpretation of $Z_c(4025)$ as the hidden charm tetraquark states via QCD Sum Rules,
  Eur.\ Phys.\ J.\ C {\bf 74}, 2810 (2014).




\bibitem{Wang:2013exa}
  Z.~G.~Wang,
 Analysis of the $Z_c(4020)$, $Z_c(4025)$, $Y(4360)$ and $Y(4660)$ as vector tetraquark states with QCD sum rules,
  Eur.\ Phys.\ J.\ C {\bf 74}, 2874 (2014).

\bibitem{Sun:2012zzd}
  Z.~F.~Sun, Z.~G.~Luo, J.~He, X.~Liu and S.~L.~Zhu,
  A note on the $B^\ast$ $\bar B$, $B^\ast\bar B^\ast$, $D^\ast\bar D$, $D^\ast\bar D^\ast$ molecular states,
  Chin.\ Phys.\ C {\bf 36}, 194 (2012).


\bibitem{Aceti:2014uea}
  F.~Aceti, M.~Bayar, E.~Oset, A.~Martinez Torres, K.~P.~Khemchandani, J.~M.~Dias, F.~S.~Navarra and M.~Nielsen,
  Prediction of an $I=1$ $D \bar D^\ast$ state and relationship to the claimed $Z_c(3900)$, $Z_c(3885)$,
  Phys.\ Rev.\ D {\bf 90}, 016003 (2014).

\bibitem{Aceti:2014kja}
  F.~Aceti, M.~Bayar, J.~M.~Dias and E.~Oset,
  Prediction of a $Z_c(4000)$ $D^* \bar D^*$ state and relationship to the claimed $Z_c(4025)$,
  Eur.\ Phys.\ J.\ A {\bf 50}, 103 (2014).

\bibitem{He:2015mja}
  J.~He,
The $Z_c(3900)$ as a resonance from the $D\bar{D}^*$ interaction,
  Phys.\ Rev.\ D {\bf 92}, 034004 (2015).


\bibitem{Dong:2013iqa}
  Y.~Dong, A.~Faessler, T.~Gutsche and V.~E.~Lyubovitskij,
  Strong decays of molecular states $Z_{c}^{+}$ and $Z_{c}^{\prime+}$,
  Phys.\ Rev.\ D {\bf 88}, 014030 (2013).

\bibitem{Wilbring:2013cha}
  E.~Wilbring, H.-W.~Hammer and U.-G.~Mei$\ss$ner,
Electromagnetic structure of the $Z_c(3900)$,
  Phys.\ Lett.\ B {\bf 726}, 326 (2013).


\bibitem{Khemchandani:2013iwa}
  K.~P.~Khemchandani, A.~Martinez Torres, M.~Nielsen and F.~S.~Navarra,
Relating $D^* \bar{D}^*$ currents with $J^P= 0^+,1^+$ and $2^+$ to $Z_c$ states,
  Phys.\ Rev.\ D {\bf 89}, 014029 (2014).



\bibitem{Guo:2013sya}
  F.~K.~Guo, C.~Hidalgo-Duque, J.~Nieves and M.~P.~Valderrama,
  Consequences of heavy quark symmetries for hadronic molecules,
  Phys.\ Rev.\ D {\bf 88}, 054007 (2013).

\bibitem{Ke:2013gia}
  H.~W.~Ke, Z.~T.~Wei and X.~Q.~Li,
  Is $Z_c(3900)$ a molecular state,
  Eur.\ Phys.\ J.\ C {\bf 73},  2561 (2013).


\bibitem{Chen:2013omd}
  W.~Chen, T.~G.~Steele, M.~L.~Du and S.~L.~Zhu,
$D^*\bar D^*$ molecule interpretation of $Z_c(4025)$,
  Eur.\ Phys.\ J.\ C {\bf 74}, 2773 (2014).

\bibitem{Wang:2013cya}
  Q.~Wang, C.~Hanhart and Q.~Zhao,
  Decoding the Riddle of $Y(4260)$ and $Z_c(3900)$,
  Phys.\ Rev.\ Lett.\  {\bf 111}, 132003 (2013).

\bibitem{Li:2014pfa}
  G.~Li, X.~H.~Liu and Z.~Zhou,
More hidden heavy quarkonium molecules and their discovery decay modes,
  Phys.\ Rev.\ D {\bf 90}, 054006 (2014).

\bibitem{Li:2013xia}
  G.~Li,
Hidden-charmonium decays of $Z_c(3900)$ and $Z_c(4025)$ in intermediate meson loops model,
  Eur.\ Phys.\ J.\ C {\bf 73}, 2621 (2013).







\bibitem{Swanson:2014tra}
  E.~S.~Swanson,
$Z_b$ and $Z_c$ exotic states as coupled channel cusps,
  Phys.\ Rev.\ D {\bf 91}, 034009 (2015).


\bibitem{Ikeda:2016zwx}
  Y.~Ikeda, S. Aoki, T. Doi, S. Gongyo, T. Hatsuda, T. Inoue, T. Iritani, N. Ishii, K. Murano, and K. Sasaki (HAL QCD Collaboration), Fate of the Tetraquark Candidate $Z_c$(3900) from Lattice QCD,
  Phys.\ Rev.\ Lett.\  {\bf 117}, 242001 (2016).


\bibitem{Voloshin:2013dpa}
  M. B.~Voloshin,
  $Z_c(3900)$$-$what is inside?,
   Phys.\ Rev.\ D {\bf 87}, 091501 (2013).

\bibitem{Chen:2011xk}
  D.~Y.~Chen and X.~Liu,
  Predicted charged charmonium-like structures in the hidden-charm dipion decay of higher charmonia,
  Phys.\ Rev.\ D {\bf 84}, 034032 (2011).

\bibitem{Chen:2013coa}
  D.~Y.~Chen, X.~Liu and T.~Matsuki,
  Reproducing the $Z_c(3900)$ structure through the initial-single-pion-emission mechanism,
  Phys.\ Rev.\ D {\bf 88}, 036008 (2013).



\bibitem{Chen:2013wca}
  D.~Y.~Chen, X.~Liu and T.~Matsuki,
  Predictions of Charged Charmoniumlike Structures with Hidden-Charm and Open-Strange Channels,
  Phys.\ Rev.\ Lett.\  {\bf 110}, 232001 (2013).

\bibitem{Liu:2013vfa}
  X.~H.~Liu and G.~Li,
  Exploring the threshold behavior and implications on the nature of $Y(4260)$ and $Z_c(3900)$,
  Phys.\ Rev.\ D {\bf 88}, 014013 (2013).








\bibitem{Yuan:2018inv}
  C.~Z.~Yuan,
  The XYZ states revisited,
  Int.\ J.\ Mod.\ Phys.\ A {\bf 33}, 1830018 (2018).

\bibitem{Meng:2007tk}
  C.~Meng and K.~T.~Chao,
  Scalar resonance contributions to the dipion transition rates of $\Upsilon(4S,5S)$ in the re-scattering model,
  Phys.\ Rev.\ D {\bf 77}, 074003 (2008).

\bibitem{Meng:2008bq}
  C.~Meng and K.~T.~Chao,
 $\Upsilon(4S,5S)\to \Upsilon(1S) \eta$ transitions in the rescattering model and the new {\it BABAR} measurement,
  Phys.\ Rev.\ D {\bf 78}, 074001 (2008).




\bibitem{Casalbuoni:1996pg}
  R.~Casalbuoni, A.~Deandrea, N.~Di Bartolomeo, R.~Gatto, F.~Feruglio and G.~Nardulli,
  Phenomenology of heavy meson chiral Lagrangians,
  Phys.\ Rep.\  {\bf 281}, 145 (1997).

\bibitem{Colangelo:2003sa}
  P.~Colangelo, F.~De Fazio and T.~N.~Pham,
  Nonfactorizable contributions in $B$ decays to charmonium: The case of $B^-\to K^- h_c$,
  Phys.\ Rev.\ D {\bf 69}, 054023 (2004).

\bibitem{Chen:2015igx}
  D.~Y.~Chen and Y.~B.~Dong,
  Radiative decays of the neutral $Z_c(3900)$,
  Phys.\ Rev.\ D {\bf 93}, 014003 (2016).

\bibitem{Chen:2014sra}
  D.~Y.~Chen, X.~Liu and T.~Matsuki,
  Observation of $e^+e^-\to \chi_{c0}\omega$ and missing higher charmonium $\psi(4S)$,
  Phys.\ Rev.\ D {\bf 91}, 094023 (2015).

\bibitem{Xiao:2017uve}
  C.~J.~Xiao and D.~Y.~Chen,
  Analysis of the hidden bottom decays of $Z_b(10610)$ and $Z_b(10650)$ via final state interaction,
  Phys.\ Rev.\ D {\bf 96}, 014035 (2017).
  
  
\bibitem{Tanabashi:2018oca}
  M.~Tanabashi {\it et al.} [Particle Data Group],
  Review of particle physics,
  Phys.\ Rev.\ D {\bf 98}, 030001 (2018).
  
\bibitem{Gortchakov:1995im}
  O.~Gortchakov, M.~P.~Locher, V.~E.~Markushin and S.~von Rotz,
  Two meson doorway calculation for $\bar p p\to \phi \pi$ including off-shell effects and the OZI rule,
  Z.\ Phys.\ A {\bf 353}, 447 (1996).
  
  
  
\bibitem{Cheng:2004ru}
  H.~Y.~Cheng, C.~K.~Chua and A.~Soni,
Final state interactions in hadronic $B$ decays,
  Phys.\ Rev.\ D {\bf 71}, 014030 (2005).



\bibitem{Lin:1999ad}
  Z.~w.~Lin and C.~M.~Ko,
  A model for $J /\psi$ absorption in hadronic matter,
  Phys.\ Rev.\ C {\bf 62}, 034903 (2000).


\bibitem{Oh:2000qr}
  Y.~s.~Oh, T.~Song and S.~H.~Lee,
  $J /\psi$ absorption by $\pi$ and $\rho$ mesons in meson exchange model with anomalous parity interactions,
  Phys.\ Rev.\ C {\bf 63}, 034901 (2001).



\bibitem{Isola:2003fh}
  C.~Isola, M.~Ladisa, G.~Nardulli and P.~Santorelli,
  Charming penguins in $B\to K^* \pi$, $K(\rho,\ \omega,\ \phi)$ decays,
  Phys.\ Rev.\ D {\bf 68}, 114001 (2003).

\bibitem{Chen:2010re}
  D.~Y.~Chen, Y.~B.~Dong and X.~Liu,
  Long-distant contribution and $\chi_{c1}$ radiative decays to light vector meson,
  Eur.\ Phys.\ J.\ C {\bf 70}, 177 (2010).

  




\end{thebibliography}
\end{document}